\documentclass[manuscript,english]{revtex4}
\usepackage[T1]{fontenc}
\usepackage[latin1]{inputenc}
\usepackage{longtable}
\usepackage{graphicx}

\makeatletter

\providecommand{\tabularnewline}{\\}

\usepackage{babel}
\makeatother
\begin{document}

\title{Compositional disorder and its influence on the structural, electronic
and magnetic properties of $MgC(Ni_{1-x}Co_{x})_{3}$ alloys using
first-principles }

\author{P. Jiji Thomas Joseph and Prabhakar P. Singh}

\affiliation{Department of Physics, Indian Institute of Technology Bombay, Mumbai 400076, India}

\begin{abstract}
First-principles, density-functional based electronic structure calculations
are carried out for $MgC(Ni_{1-x}Co_{x})_{3}$ alloys over the concentration
range $0\leq x\leq1$, using Korringa-Kohn-Rostoker coherent-potential
approximation (KKR CPA) method in the atomic sphere approximation
(ASA). The self-consistent calculations are used to study the changes
as a function of $x$ in the equation of state parameters, total and
partial densities of states, magnetic moment and the on-site exchange
interaction parameter. To study the magnetic properties as well as
its volume dependence, fixed-spin moment calculations in conjunction
with the phenomenological Landau theory are employed. The salient
features that emerge from these calculations are (i) a concentration
independent variation in the lattice parameter and bulk modulus at
$x$$\sim$$0.75$ with an anomaly in the variation of the pressure
derivative of bulk modulus, (ii) the fixed-spin moment based corrections
to the overestimated magnetic ground state for $0.0\leq x\leq0.3$
alloys, making the results consistent with the experiments, and (iii)
the possibility of multiple magnetic states at $x$ $\sim$$0.75$,
which, however,  requires further improvements in the calculations.
\end{abstract}
\maketitle

\section{Introduction}

Much of the interest in the $8$K cubic perovskite superconductor
$MgCNi_{3}$ \cite{nature-411-54} stems from its spectral distribution
of electronic states \cite{prl-88-027001,prb-64-140507,prb-64-180510,prb-64-100508,prb-66-174512,prb-66-172507}.
A sharp peak, primarily composed of $Ni$ $3d$ bands show up in the
density of states spectra, suggesting that the material is close to
a ferromagnetic instability \cite{prl-88-027001,prb-64-140507}. Theoretical
calculations based on the Stoner model as well as the NMR experiments
reveal the possibility of co-existence of spin-fluctuation with superconductivity
\cite{prb-64-140507,prl-87-257601}. This makes $MgCNi_{3}$ a material
in a class of its own, since according to conventional models of superconductivity
magnetic scattering is thought to induce pair-breaking effects. A
simple case study is that of the binary cubic $VN$ alloy in which
the expected superconducting transition temperature is about $30$K,
which is lowered down to $\sim$$6$K due to spin-fluctuations \cite{prb-22-4284,prb-24-155}. 

The origin of the incipient magnetism in $MgCNi_{3}$ could be partly
attributed to the presence of $C$ in its octahedral interstitial
position \cite{condmat-0504659}. The spatially extended $C$ $2p$
orbitals strongly hybridize with that of the $Ni$ $3d$, de-localizing
the electronic states. As a result the density of states at the Fermi
energy, $N(E_{F})$, falls short of satisfying the Stoner criteria.
It is suggested that approximately $0.5$ holes, which may correspond
to $10$\% of $Mg$ replaced by an alkali metal, or about $7$\% replacement
of $Ni$ with $Co$, would drive the corresponding material magnetic
\cite{prl-88-027001,prb-64-140507,prb-64-100508}. The latter follows
the rigid-band model, where the Fermi level is expected to lower in
energy with respect to hole concentration, thus making $E_{F}$ coincide
with the transition-metal-derived $3d$ singularity. 

Previously, studies of hole dopings via vacancies and $B$ substitutions
in the $C$ sub-lattice \cite{condmat-0412551} and $Fe$ and $Co$
substitutions in the $Ni$ sub-lattice have been carried out \cite{prb-66-024520,jap-91-8504,prb-65-064525,ssc-119-491,condmat-0105366,prb-66-064510,sst-15-1316,prb-68-064503,bjp-32-755,sst-17-274}.
In none of the above cases, a definite magnetic solution was found
feasible. For $Co$ substitutions the experimental results are unambiguous
over the fact that $T_{C}$ decreases as $Co$ $at$\% increases in
the disordered $MgC(Ni_{1-x}Co_{x})_{3}$ alloys \cite{prb-66-024520,prb-65-064525,condmat-0105366,prb-66-064510}.
However, they remain controversial on the rate at which the $T_{C}$
is decreased \cite{condmat-0105366,prb-66-064510}. Resistivity characterization
shows that $T_{C}$ decreases gradually with increasing $x$. A small
superconducting transition is observed for the samples with $x$=$0.5$,
which indicates that the superconducting volume fraction decreases
upon substitution of $Co$ for $Ni$ \cite{prb-66-064510}. However,
this is inconsistent with the susceptibility measurements where $Co$
in dilute limit is found to completely suppress the superconductivity,
although no magnetic long-range order is detected \cite{ssc-119-491}.
The rapid loss of superconductivity is argued to be consistent with
magnetic quenching of the superconductivity, probably in the form
of spin-fluctuations. 

$Co$ forms a complete solid solution in $MgCNi_{3}$ in the cubic
perovskite structure. In $MgC(Ni_{1-x}Co_{x})_{3}$ alloys, the ferromagnetic
onset is estimated to be around $0.75$ $at$\% of $Co$ \cite{condmat-0105366}.
Further, experiments as well as theoretical calculations suggest that
a complete replacement of $Ni$ with $Co$ drive in a definite ferromagnetic
ground state \cite{prb-66-024520,jap-91-8504,prb-65-064525}. The
$Co$ local moment in $MgCCo_{3}$ is calculated to be $0.33$$\mu_{B}$
with the magnetization energy being equal to $22$mRy \cite{jap-91-8504,prb-65-064525}.
Theoretical calculations also reveal that the stoichiometrically ordered
alloys $MgCNi_{2}Co$ and $MgCNiCo_{2}$ remain non-magnetic \cite{jap-91-8504,prb-65-064525}. 

As a precursor to understanding the nature of magnetic instability
in $MgCNi_{3}$, a theoretical study of $Co$ substitution in the
$Ni$ sub-lattice was carried out. To study the effects of compositional
disorder on the structural, electronic and magnetic properties of
$MgC(Ni_{1-x}Co_{x})_{3}$ alloys we have employed the Green's function
technique in conjunction with the coherent-potential approximation
\cite{progmatsci-1,kluwer-1997}. The versatility of the method allows
a relative comparison of the alloy energetics of the stoichiometrically
ordered alloys and their disordered counterparts. 

In the present work, we have studied the changes in the equation of
state parameters, density of states and the magnetic properties as
a function of $x$ in $MgC(Ni_{1-x}Co_{x})_{3}$ alloys. Both self-consistent
as well as the fixed-spin moment method \cite{jphysf-14-l129} in
conjunction with the phenomenological Landau theory of phase transition
are employed to study the magnetic properties of the disordered $MgC(Ni_{1-x}Co_{x})_{3}$
alloys. 

The paper is organized as follows. In Sec. 2, we briefly describe
the computational details. Sec. 3 contains the results and discussion
in terms of equation of state parameters, electronic structure and
magnetic properties of $MgC(Ni_{1-x}Co_{x})_{3}$ alloys.

\section{Computational details}

The ground state properties are calculated using the Korringa-Kohn-Rostoker
(KKR) method formulated in the atomic-sphere approximation (ASA) \cite{progmatsci-1,kluwer-1997}.
Chemical disorder in the $Ni$ sub-lattice with $Co$, are assumed
to be completely random and is accounted for by means of coherent-potential
approximation (CPA) \cite{pr-156-809}. For improving the alloy energetics,
the ASA is corrected by the use of both the muffin-tin correction
for the Madelung energy \cite{prl-55-600} and the multi-pole moment
correction to the Madelung potential and energy \cite{prb-66-024201,prb-66-024202}.
These corrections bring significant improvements in the calculations
by taking into account the non-spherical part of the polarization
effects \cite{cms-15-119}. Note that the electrostatic interactions
are obviously a key contribution to the effective interactions between
differently charged cations.

The partial waves in the KKR-ASA calculations are expanded up to $l_{max}$=$3$
inside the atomic spheres. The multi-pole moments of the electron
density have been determined up to $l_{max}^{M}$=$6$, and then used
for the multi-pole moment correction to the Madelung energy. The exchange-correlation
effects are taken into consideration via the local-density approximation
(LDA) with Perdew and Wang parametrization \cite{prb-45-013244}.
The core states have been recalculated after each iteration. The calculations
are partially scalar relativistic in the sense that although the wave
functions are non-relativistic, first order perturbation corrections
to the energy eigenvalues due to the Darwin and the mass-velocity
terms are included. Further, screening constants $\alpha$ and $\beta$
were incorporated in the calculations, following the prescription
of Ruban and Skriver \cite{prb-66-024201}. These values were estimated
from the order($N$) locally self-consistent Green's function method
\cite{prb-56-9319} and were determined to be $0.83$ and $1.18$,
respectively. The atomic sphere radii of $Mg$, $C$ and $Ni$/$Co$
were kept as $1.404$, $0.747$, and $0.849$ of the Wigner-Seitz
radius, respectively. The overlap volume of the atomic spheres was
less than $15$\%, which is legitimate within the accuracy of the
approximation \cite{lmto-1984}. The total energies were calculated
with $1771$ $\mathbf{k}$-points in the irreducible wedge of the
Brillouin zone. The convergence in charge density was achieved so
that the root-mean square of moments of the occupied partial density
of states becomes smaller than $10^{-6}$. 

Numerical calculations of magnetic energy $\Delta E(M)$ for $MgC(Ni_{1-x}Co_{x})_{3}$
alloys were carried out at their self-consistently determined equilibrium
lattice constants using the fixed-spin moment method of alloy theory
\cite{jphysf-14-l129}. In the fixed-spin moment method the total
energy is obtained for a given magnetization $M$, i.e., by fixing
the numbers of electrons with up and down spins. In this case, the
Fermi energies in the up and down spin bands are not equal to each
other because the equilibrium condition would not be satisfied for
arbitrary $M$. At the equilibrium $M$ two Fermi energies will coincide
with each other. The total magnetic energy becomes minimum or maximum
at this value of $M$. 

The diffraction experiments measure displacements as large as $0.05\textrm{Å}$
for the $Ni$ atoms in $MgCNi_{3}$. These, however are found to be
less significant towards any qualitative analysis of the materials
properties. The first-principles calculations ignoring these effects,
thereby assuming a rigid perfect underlying cubic symmetry, find no
profound magnetic effects emerging, unlike in the case of $Fe$ impurities
in FCC $Al$ or in BCC $Zr$ \cite{prb-69-165116,prb-57-7004,prl-67-3832,cms-8-131,prl-71-4206,prl-71-3525}.
Although in disordered alloys lattice relaxations around an impurity
is important, for $3d$ impurities in a $3d$ host these effects seem
to be less important \cite{cms-8-131,prb-55-4157,prb-39-930}. Usually,
such effects stem from large atomic size mismatch of the host and
the impurity atoms which is, however, not the case for $Co$ and $Ni$
ions.

\section{Results and discussion}

The $MgC(Ni_{1-x}Co_{x})_{3}$ alloys are theoretically characterized
in terms of the variation of physical properties as a function of
$x$. These include the equation of state parameters such as lattice
constant, bulk modulus and its pressure derivative, electronic structure
expressed in terms of total, sub-lattice resolved partial and $l-$
decomposed density of states. The magnetic properties are determined
via both the self-consistent calculations and the fixed-spin moment
calculations, including the variation in the $Co$ local moment and
the on-site exchange interaction constant as a function of $x$. For
$x<0.3$, the self-consistent calculations yield ambiguous results
about the magnetic ground state of $MgC(Ni_{1-x}Co_{x})_{3}$ alloys,
therefore, we used the fixed-spin moment method in this concentration
range. These calculations find that for low $Co$ rich alloys, the
ground state is definitely paramagnetic. In conjunction with the Landau
theory the $MgC(Ni_{1-x}Co_{x})_{3}$ alloys for $0.0\leq x\leq0.3$
show a greater propensity of magnetism. However, both the self-consistent
as well as the fixed-spin moment calculations yield consistent results
for $x\geq0.75$ alloys.

\subsection{Equation of state parameters}

The equilibrium lattice constants of $MgC(Ni_{1-x}Co_{x})_{3}$ alloys
for $0\leq x\leq1$ were determined by calculating the total energy
at six lattice constants close to the expected equilibrium lattice
constant and then using the third-order Birch-Murnaghan equation of
state \cite{jgr-57-227,wiley-140}. Since the choice of the exchange
correlation term in the Kohn-Sham Hamiltonian is crucial in determining
the equation of state parameters we have carried out the calculations
using three different exchange correlation functionals namely the
local density approximation (LDA), the generalized gradient approximation
(GGA) \cite{prb-54-16533}, and the local Airy gas (LAG) approximation
\cite{prb-62-10046}. One may note that the Birch- Murnaghan equation
is derived from the theory of finite strain, by considering an elastic
isotropic medium under isothermal compression, with the assumption
that the pressure- volume relation remains linear. The polynomial
would yield a reasonable guess to the equation of state parameters
provided the fit is chosen in a volume range close to equilibrium. 

The equilibrium lattice constants for $MgC(Ni_{1-x}Co_{x})_{3}$ alloys
decrease as $x$ increases. This is consistent with the fact that
the ionic radii of $Co$ is less than that of $Ni$. For $MgCNi_{3}$
the equilibrium lattice constant was calculated to be $7.139$, $7.305$
and $7.233$ $a.u.$, in the LDA, GGA and LAG approximations, respectively.
When compared to the experiments, we find that LDA underestimates
the value of the lattice constant by $1$\%, while both GGA and LAG
overestimate the value. The LDA estimate matches well with the previous
full-potential report of Singh and Mazin \cite{prb-64-140507}. We
note that the consistency with the full-potential methods owes to
the muffin-tin correction to the ASA, without which the KKR-ASA calculations
find the equilibrium lattice constant for $MgCNi_{3}$ as $6.986$
a.u. For $MgCCo_{3}$, the equilibrium lattice constants for both
spin polarized and unpolarized calculations turn out to be 7.078 and
7.071 $a.u$, respectively. A small increase in the lattice constant
in the spin polarized calculations is consistent with the fact that
magnetization through exchange splitting requires larger volume.

\begin{figure}[t]
~

\includegraphics[%
  scale=0.3]{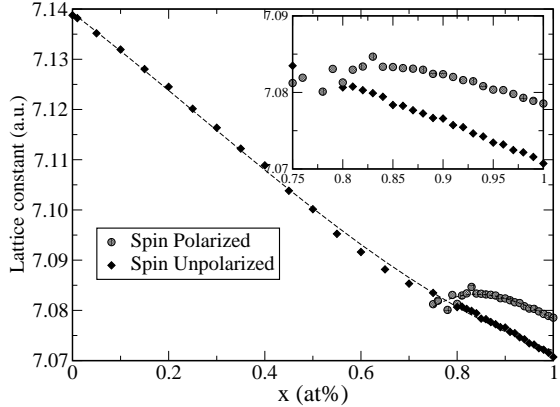}

\caption{\label{fig-co3LatConc}The variation in the lattice constant of $MgC(Ni_{1-x}Co_{x})_{3}$
as a function of $x$ calculated using the KKR-ASA-CPA method. Results
for both spin polarized and unpolarized calculations are shown. }
\end{figure}

\begin{figure}[ht]
~

\includegraphics[%
  scale=0.3]{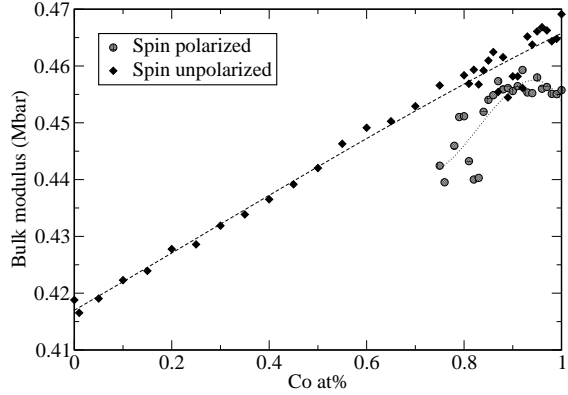}

\caption{\label{fig-co3BlkConc}The variation in the bulk modulus for $MgC(Ni_{1-x}Co_{x})_{3}$
alloys as a function of $x$, calculated using the KKR-ASA-CPA 
method. Results for both spin polarized and unpolarized
calculations are shown. }
\end{figure}

The variation of the lattice constants as a function of $x$ in $MgC(Ni_{1-x}Co_{x})_{3}$
alloys, following both spin unpolarized and polarized calculations
is shown in Fig.\ref{fig-co3LatConc}. The salient features that emerge
are (i) deviation from the Vegard's law, (ii) a  weak $x$ dependence 
of the lattice constant in the range $0.7\leq x\leq0.8$,
and (iii) deviation of lattice constants above $x=0.7$. Deviation
from the Vegard's law suggests a diversity in the inter-atomic interactions.
The Vegard's law implicitly assumes that the chemical constituents
which make up the alloy have similar potentials and that their distribution
remains truly random. Although one may assume $Co$ and $Ni$ to have
similar features, the nature of interactions in them is significantly
different because $MgCNi_{3}$ is paramagnetic while $MgCCo_{3}$
is ferromagnetic. This indicates that the $C$ $2p$ hybridization
with the transition metal $3d$ bands in $MgCNi_{3}$ and $MgCCo_{3}$
are different and that the $Ni$-$C$ hybridization would be stronger
than the $Co$-$C$ counterpart in $MgCCo_{3}$. 

The insensitivity of the lattice constant of $MgC(Ni_{1-x}Co_{x})_{3}$
alloys with respect to $x$ in the range $0.7<x<0.8$ arises due to
a balance between the chemical binding forces and the magnetic ones.
Chemical binding tends to compress the lattice while magnetization
via exchange splitting of bands requires large volume. These two opposing
forces set in to provide an invariance in the lattice constant for
$0.7<x<0.8$ in $MgC(Ni_{1-x}Co_{x})_{3}$ alloys. For $x>0.7$, the
spin polarized calculations yield slightly higher lattice constants
compared to the spin unpolarized calculations, which is purely due
to the magnetic effects. 

The bulk modulus is proportional to the second derivative of the energy-
volume curve. The LDA usually underestimates the lattice constant
thus leading to an overestimation of bulk modulus. The emphasis here,
however, is to make a qualitative judgment about the bulk modulus
as a function of $x$. Fig.\ref{fig-co3BlkConc} shows the change
in the isothermal bulk modulus of $MgC(Ni_{1-x}Co_{x})_{3}$ alloys
as a function of $x$ for $0\leq x\leq1$. We find that the bulk modulus
increases with increasing $x$. 

Assuming that the alloys are isotropic and that the Debye approximation
holds, the bulk modulus scales in proportion to the Debye temperature
$\Theta_{D}$ ($\propto$$B_{0}^{\frac{1}{2}}$) . Higher the value
of $\Theta_{D}$ higher the lattice stiffness. In $MgCNi_{3}$ the
frequency of a soft acoustic $Ni$ based phonon mode calculated in
the harmonic approximation tends to become negative \cite{prb-68-220504,prb-69-092511}.
Stabilization of this mode results in a dynamic displacement of the
$Ni$ ions perpendicular to the $Ni$-$C$ direction which then allows
each of the $Ni$ atoms to move away from the neighboring $C$ atoms
towards the empty interstitial. Partial replacement of $Ni$ with
$Co$ weakens the bonding with the $C$ atoms. This reduces the advantage
of relaxations or distortions thereby increasing $\Theta_{D}$, thus
a concurrent hardening of the phonon modes. When compared to the spin
unpolarized calculations, the spin polarized ones yield a lower bulk
modulus, which is directly correlated with their corresponding lattice
constants.

\begin{figure}[t]
~

\includegraphics[%
  scale=0.3]{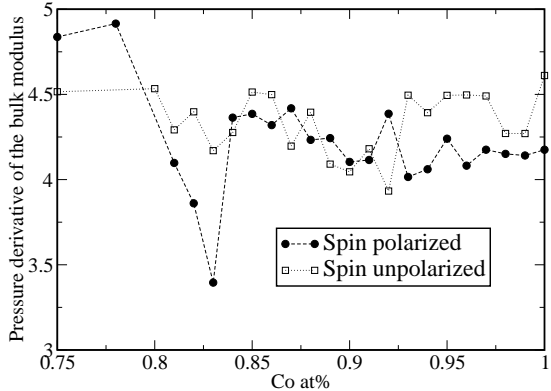}

\caption{\label{fig-co3-prBlkConc}The change in the pressure derivative of
bulk modulus of $MgC(Ni_{1-x}Co_{x})_{3}$ as a function of $x$ calculated
using the KKR-ASA-CPA method. Results for both spin polarized
and unpolarized calculations are shown. }
\end{figure}

Since the pressure derivatives of the bulk modulus involves higher
order derivatives, they are often noisy. Qualitatively, however, in
the Debye approximation the pressure derivatives of the bulk modulus
contain the information of the averaged lattice vibrations of the
material via the Gruneissen parameter. If one assumes that all the
vibrational modes respond to volume in a similar fashion, then they
are often useful in interpreting the structural details of the material.
The variation in the pressure derivative of the bulk modulus, as shown
in Fig.\ref{fig-co3-prBlkConc}, suggests an anomaly at $x=0.82$.
The discontinuity in the case of spin polarized results suggests that
the anomaly could be related to magnetic effects because it appears
in the proximity of magnetic onset in the $MgC(Ni_{1-x}Co_{x})_{3}$
alloys. 

\begin{longtable}[!t]{|c||c|c|c|}
\hline 
&
LDA (Ry)&
GGA (Ry)&
LAG (Ry)\tabularnewline
\hline
\hline 
$MgCNi_{2}Co$&
-1866.120100&
-1866.883569&
-1869.423046\tabularnewline
\hline 
$MgC(Ni_{0.66}Co_{0.33})_{3}$&
-1866.118525&
-1866.882022&
-1869.421520\tabularnewline
\hline
\hline 
\textbf{$\Delta E$ (mRy)}&
1.575&
1.547&
1.526\tabularnewline
\hline
\hline 
$MgCNiCo_{2}$&
-1815.219322&
-1815.974307&
-1818.477701\tabularnewline
\hline 
$MgC(Ni_{0.33}Co_{0.66})_{3}$ &
-1815.219848&
-1815.974861&
-1818.478283\tabularnewline
\hline
\hline 
\textbf{$\Delta E$ (mRy)}&
-0.526&
-0.554&
-0.573\tabularnewline
\hline
\caption{\label{tab-stodis-Enr}Comparison of the equilibrium total energy
expressed in Ry units for $x=\frac{1}{3}$ and $x=\frac{2}{3}$ alloys
of $MgC(Ni_{1-x}Co_{x})_{3}$ in three different approximations to
$V_{xc}$; LDA, GGA and LAG and their difference ($\Delta E=E_{ordered}-E_{disordered}$)}


\end{longtable}

The versatility of the KKR-ASA method is that it allows a direct comparison
of the structural energies of both disordered and ordered alloys.
Two ordered alloys can be identified in the concentration range $0<x<1$.
These are $MgCNi_{2}Co$ and $MgCNiCo_{2}$, which correspond to $x=0.33$
and $0.66$ respectively in the case of disordered $MgC(Ni_{1-x}Co_{x})_{3}$
alloys. Either of these alloys falls in the $0.3<x<0.7$ range, where
both spin polarized and unpolarized calculations for the disordered
alloys yield degenerate total energies. Table.\ref{tab-stodis-Enr}
compares the KKR-ASA structural energies for the ordered and disordered
alloys, calculated using three different exchange correlation functionals
to Kohn-Sham Hamiltonian. Evidently, from Table.\ref{tab-stodis-Enr},
it follows that for $x=\frac{1}{3}$ the ordered phases are more stable
than their disordered counterparts. However, as $x$ increases to
$\frac{2}{3}$, the disordered alloys become more stable. The changes
in the structural energies provide some clues about the chemical ordering
of atoms due to magnetic effects. We plan to report the results in
a future work.

The change in the equation of state parameters upon chemical ordering
was also calculated using the KKR-ASA method in conjunction with the
Birch-Murnaghan equation of state with LDA. The results are shown
in Table.\ref{tab-co3StoDis}. The calculations shows that the disordered
alloys have slightly higher lattice constant than their stoichiometrically
ordered counterparts. 
%
 \begin{longtable}[!t]{|c|c|c|c|}
\hline 
&
$a_{eq}$($a.u.$)&
$B_{eq}$($Mbar$)&
$B_{eq}^{'}$\tabularnewline
\hline
\hline 
$MgCNi_{2}Co$&
7.113&
0.432&
4.325\tabularnewline
\hline 
$MgC(Ni_{0.66}Co_{0.33})_{3}$&
7.114&
0.433&
4.326\tabularnewline
\hline
\hline 
$MgCNiCo_{2}$&
7.086&
0.451&
4.463\tabularnewline
\hline 
$MgC(Ni_{0.33}Co_{0.66})_{3}$ &
7.088&
0.451&
4.237\tabularnewline
\hline

\caption{\label{tab-co3StoDis}Comparison of the equation of state parameters
for stoichiometrically ordered and disordered alloys using the LDA-
KKR-ASA- CPA method. }
\end{longtable}

\subsection{Electronic structure}

Electronic structure of $MgC(Ni_{1-x}Co_{x})_{3}$ alloys are characterized
both by means of spin unpolarized and polarized density of states,
calculated using the KKR-ASA-CPA method. For $x<0.75$, spin unpolarized
density of states are shown while for $x>0.75$, both spin unpolarized
and polarized density of states are shown.

\subsubsection{spin unpolarised density of states}

\begin{figure}[ht]
~

~

~

\includegraphics[%
  scale=0.3]{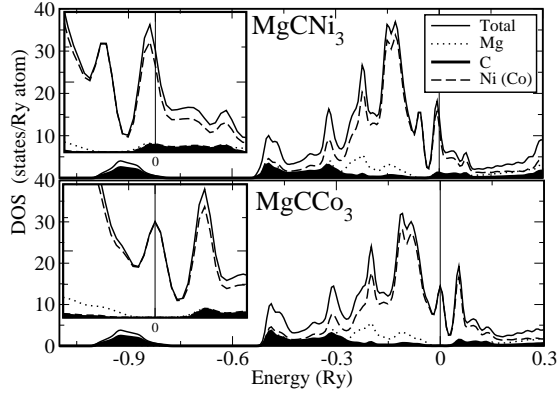}

\caption{\label{mgcni3co3DOS} The spin unpolarized total and sub-lattice
resolved partial density of states for $MgCNi_{3}$ (upper panel)
and $MgCCo_{3}$ (lower panel) calculated at their equilibrium lattice
constants. In the inset, a blow up around the Fermi energy for the
corresponding alloys are shown. The vertical line through energy zero
represents the alloy Fermi energy. }
\end{figure}

In Fig.\ref{mgcni3co3DOS}, we show the total and sub-lattice resolved
density of states of $MgCNi_{3}$ and $MgCCo_{3}$ calculated at their
equilibrium lattice constants. The isolated deep valence states in
Fig.\ref{mgcni3co3DOS} are essentially the $C$ $2s$ states. Further
higher in energy, a gap opens up followed by an admixture of $C$
$2p$ and metal $3d$ states. The strong hybridization is manifested
in the DOS as a wide energy spread of the $C$ $2p$ and the transition
metal $3d$ states. From Fig.\ref{mgcni3co3DOS} it follows that the
hybridization between the $C$-$Ni$ orbitals are much more stronger
than the $C$-$Co$ orbitals in their respective alloys. Hybridization
gives not only an important mixing between the states of the conduction
bands but also leads to a separation of the bonding and anti-bonding
states, creating a pseudo-gap. Existence of pseudo-gaps have been
reported in various crystalline solids, amorphous alloys\cite{JPCM-1-9685}
and also in quasi-crystals\cite{Europhys-34-207,JJAP-30-389}. Of
the two mechanisms, ionic and covalent, proposed for the formation
of pseudo-gap in alloys the latter is found to play a significant
role in the case of transition-metal-based inter-metallics. The pseudo-gap
in hexagonal close-packed transition metal rich systems are largely
attributed to metal $d$ resonance. Though the $Mg$-$Ni$($Co$)
bonding is weak, a peak just below $-0.4$Ry constitutes the bonding
between $3s$ electrons of $Mg$ and the transition metal $3d$ electrons
in their respective alloys. The striking difference in the DOS of
$MgCNi_{3}$ and $MgCCo_{3}$ is the absence of $C$ $2p$ states
at the $E_{F}$ for $MgCCo_{3}$. In $MgCCo_{3}$ the $N(E_{F})$
is entirely due to the $Co$ $3d$ states while for $MgCNi_{3}$ the
$C$ $2p$ contribution to $N(E_{F})$ is about $7$\%. 

The present KKR-ASA-CPA calculations find $N(E_{F})$ for $MgCNi_{3}$
to be $14.56$ state/Ry atom. The value
is, however, at variance with the previous reports. For example, Szajek
reports the value as $14.32$ state/Ry atom \cite{jpcm-13-L595},
Mazin and Singh report as $13.57$ states/Ry atom \cite{prb-64-140507},
Shim $et$ $al$ as $14.52$ states/Ry atom \cite{prb-64-180510},
Rosner $et$ $al$ as $13.06$ states/Ry atom \cite{prl-88-027001}.
Dugdale and Jarlborg \cite{prb-64-100508} report $N(E_{F})$ to be
$17.27$ and $9.49$ states/Ry atom \cite{prb-64-100508} for two different
band-structure methods with exchange-correlation effects considered
in the LDA and using the experimental lattice constant. These results
show that $N(E_{F})$ is indeed sensitive to the type of the electronic
structure method employed and also to the numerical values of the
parameters like that of the Wigner-Seitz radii and others. Note that
these differences are significant as they control the proximity to
magnetism in the Stoner model, as emphasized by Singh and Mazin \cite{prb-64-140507}.
For $MgCCo_{3}$ the value of $N(E_{F})$ is determined to be $14.43$
state/Ry atom, in agreement with the earlier
reports. 

\begin{figure}[ht]
~

~

~

\includegraphics[%
  scale=0.3]{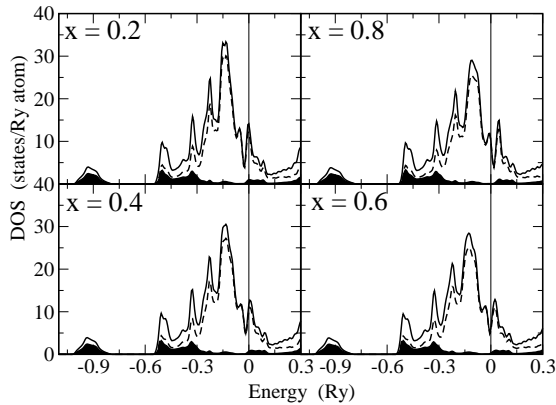}

\caption{\label{fig-co3-Totparados} The spin unpolarized total (solid line),
sub-lattice resolved $C$ (shaded) and $(Ni_{1-x}Co_{x})_{3}$
 density of states, with $x$ as shown in the panels, for
$MgC(Ni_{1-x}Co_{x})_{3}$ alloys. The vertical line through energy
zero corresponds to the alloy Fermi energy. }
\end{figure}

Fig.\ref{fig-co3-Totparados} shows the spin unpolarized, total and
sub-lattice resolved partial DOS for $MgC(Ni_{1-x}Co_{x})_{3}$ alloys
for $x$= $0.2$, $0.4$, $0.6$ and $0.8$. The overall characteristic
features of the DOS structure are preserved as $x$ increases in $MgC(Ni_{1-x}Co_{x})_{3}$,
however with the sharp structures being smeared out due to disorder.
As $x$ increases, the decreasing electron count is followed by the
inward movement of $E_{F}$ on the energy scale. The bottom of the
states, however, remain more or less fixed with respect to the alloy
$E_{F}$ as a function of $x$. At about $x=0.7$ the $E_{F}$ is
pinned in the pseudo-gap. The resulting $N(E_{F})$ is minimum for
these alloys over the whole concentration profile. As $x$ increases
further from $0.7$, the $N(E_{F})$ gradually increases due to the
$Co$ $3d$ states. A high $N(E_{F})$, predominantly metal $3d$
in character, thus increases the possibility of exchange splitting
of bands in $MgC(Ni_{1-x}Co_{x})_{3}$. 

\begin{figure}[ht]
~

~

~

\includegraphics[%
  scale=0.3]{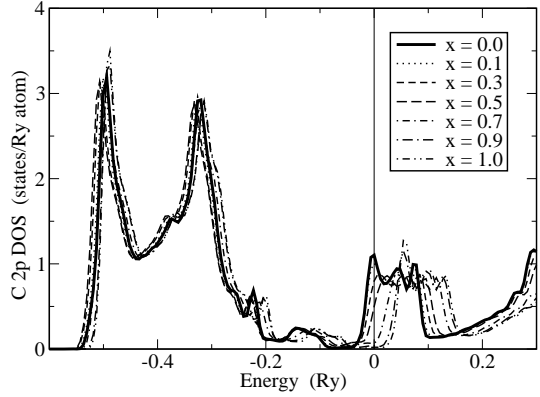}

\caption{\label{fig-co3-Cparados} The spin unpolarized sub-lattice resolved
$C$ $2p$ DOS for $MgC(Ni_{1-x}Co_{x})_{3}$ with $x$ as shown in
the panel. The vertical line through energy zero corresponds to the
alloy Fermi energy. }
\end{figure}

One may note that the $C$ $2p$ states in and around $E_{F}$ decrease
with increasing $x$, and virtually cease to exist for $x>0.7$ alloys.
Since the number of $C$ electrons remains invariant, one may then
naturally expect a significant charge redistribution on the energy
scale. This is illustrated in Fig.\ref{fig-co3-Cparados}. Though
the overall shape of the DOS remains the same, a significant charge
redistribution over the energy range $-0.6$ Ry to $-0.2$ Ry is observed. 

\begin{figure}[ht]
~

~

~

\includegraphics[%
  scale=0.3]{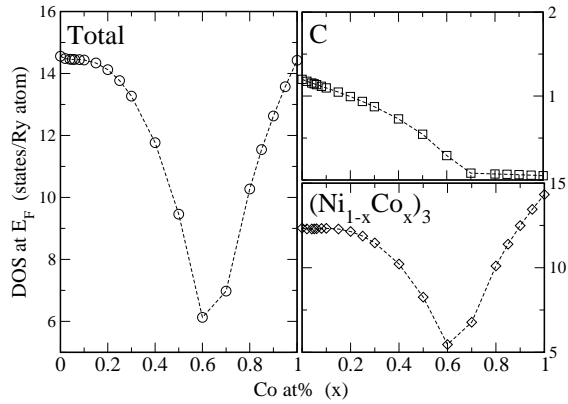}

\caption{\label{fig-co3-dosEF}The variation in the total density of states
at Fermi energy as a function of $x$ in $MgC(Ni_{1-x}Co_{x})_{3}$
for $0\leq x\leq 1$ alloys in units of states/Ry cell.}
\end{figure}

in Fig \ref{fig-co3-dosEF}, we show the evolution of the states at
$E_{F}$ as a function of $x$ in $MgC(Ni_{1-x}Co_{x})_{3}$ alloys.
Initially with increase in $x$, the $N(E_{F})$ slightly increases,
however, not as expected from the rigid band model of alloy theory.
Disorder smears out the $Ni$ $3d$- derived van Hove peak of $MgCNi_{3}$
considerably. 

Thus the essential difference between the DOS of $MgCNi_{3}$ and
$MgCCo_{3}$ is the absence of $C$ contribution to $N(E_{F})$. One
can find that for both $MgCNi_{3}$ and $MgCCo_{3}$ the $N(E_{F})$
are comparable and are equal to 14.557 states/Ry atom and 14.425 states/Ry atom,
respectively. However, their properties are found to be significantly
different. $MgCNi_{3}$ is non-magnetic with incipient magnetism akin
to spin-fluctuations being anticipated, while $MgCCo_{3}$ is a non
superconductor with a definite ferromagnetic state. 

\begin{figure}[ht]
~

~

~

\includegraphics[%
  scale=0.3]{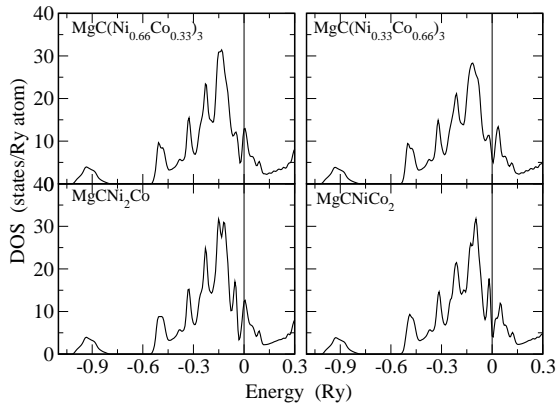}

\caption{\label{stodis-totdos}A comparison of the spin unpolarized total
DOS of stoichiometrically ordered and disordered $MgC(Ni_{1-x}Co_{x})_{3}$
alloys for $x$= $\frac{1}{3}$ and $\frac{2}{3}$, respectively. The vertical line
represents the Fermi energy. }
\end{figure}

\begin{figure}[ht]
\includegraphics[%
  scale=0.3]{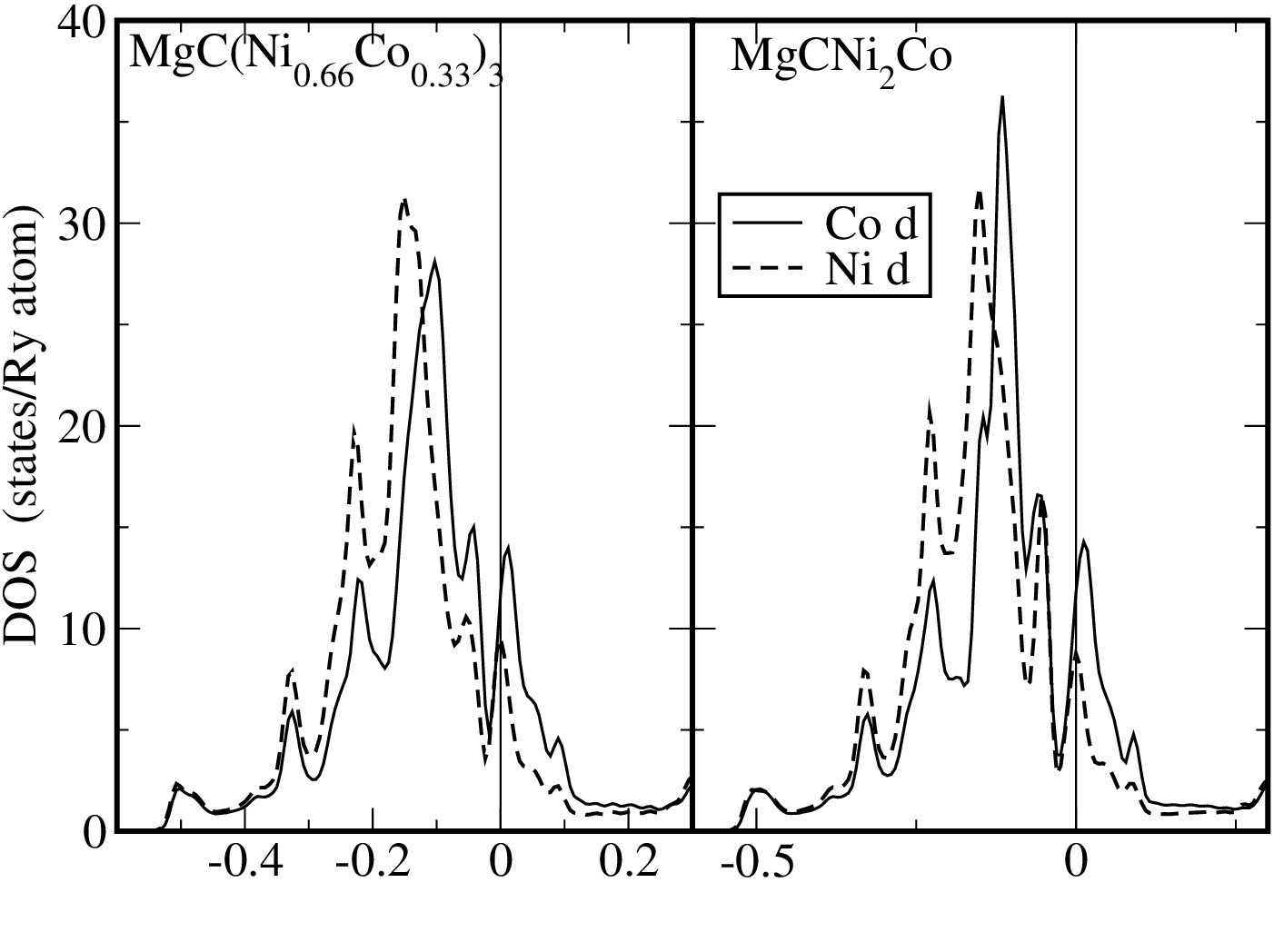}

\includegraphics[%
  scale=0.3]{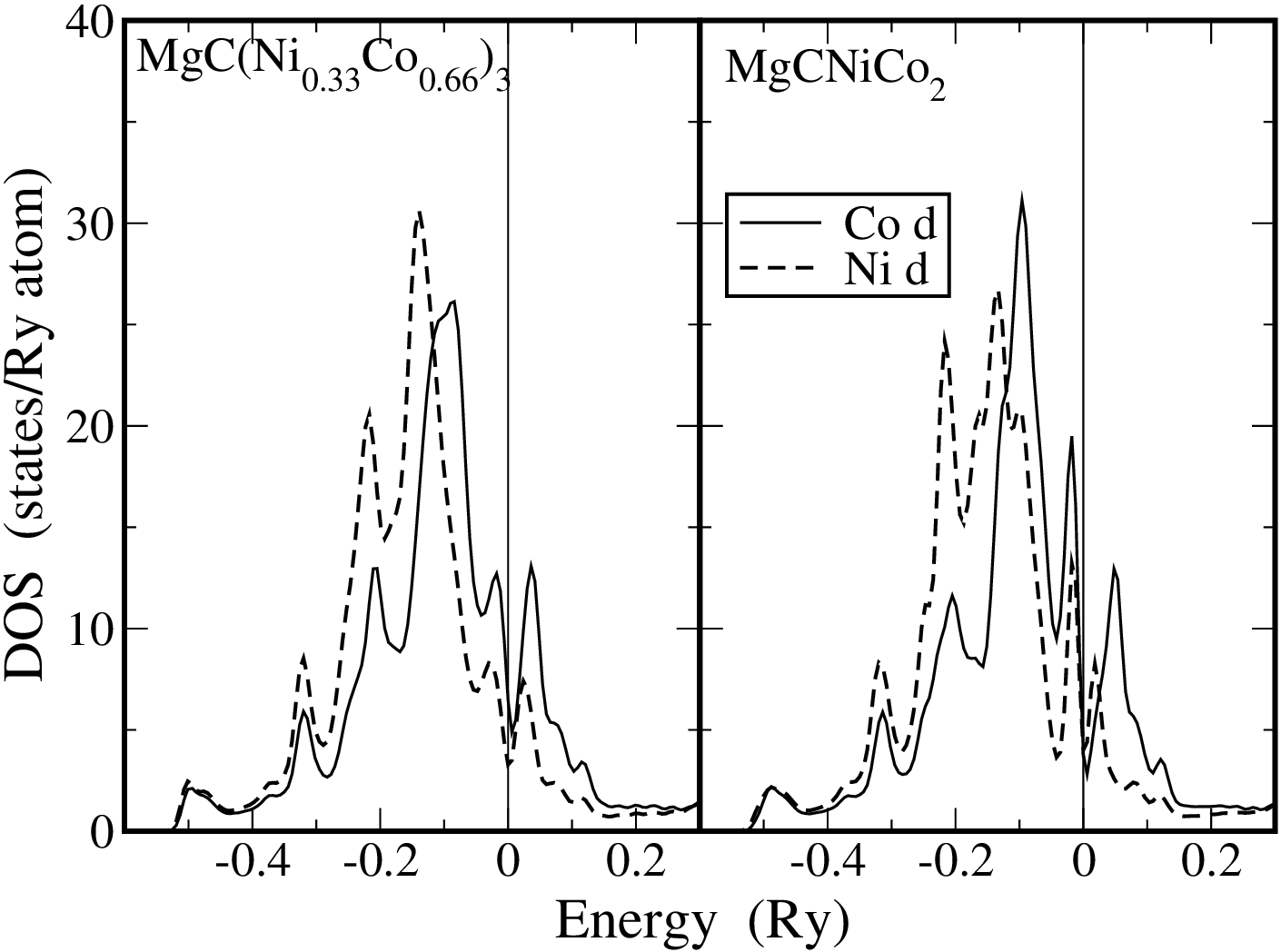}

\caption{\label{stodis-NiCodos}A comparison of spin unpolarized, sub-lattice
resolved transition metal DOS of stoichiometrically ordered and disordered
$MgC(Ni_{1-x}Co_{x})_{3}$ alloys for $x$= $\frac{1}{3}$ (upper
panel) and $\frac{2}{3}$ (lower panel). The vertical line represents
the Fermi energy. }
\end{figure}

\begin{figure}[ht]
~

~

\includegraphics[%
  scale=0.3]{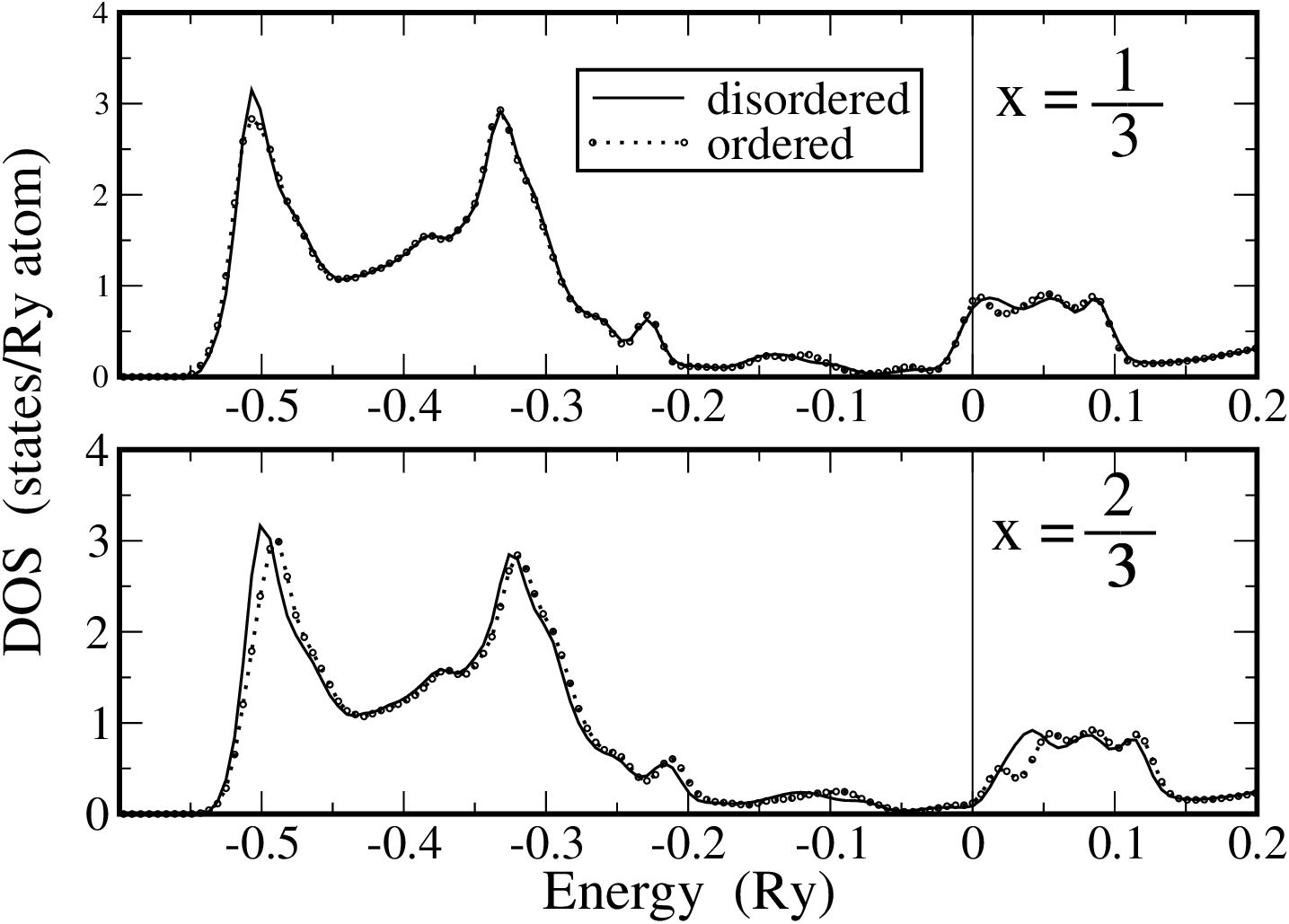}

\caption{\label{stodis-C-dos}A comparison of spin unpolarized sub-lattice
resolved partial $C$ $2p$ DOS of stoichiometrically ordered and
disordered $MgC(Ni_{1-x}Co_{x})_{3}$ alloys for $x$= $\frac{1}{3}$
(upper panel) and $\frac{2}{3}$ (lower panel). The vertical line
represents the Fermi energy. }
\end{figure}

The change in the total density of states for the chemically ordered
$MgCNi_{2}Co$ and $MgCNiCo_{2}$ alloys with respect to their disordered
counterparts, $MgC(Ni_{0.66}Co_{0.33})_{3}$ and $MgC(Ni_{0.33}Co_{0.66})_{3}$
respectively, are shown in Fig.\ref{stodis-totdos}. The overall characteristic
features of the DOS remain essentially the same, with exception to
the peak broadening in the disordered alloys. The change in the $Ni$
$3d$ and $Co$ $3d$ states for the corresponding ordered and disordered
alloys are shown in Fig.\ref{stodis-NiCodos} and the change in the
$C$ partial DOS is shown in Fig.\ref{stodis-C-dos}. 

\subsubsection{Spin polarized density of states}

With $N(E_{F})$ in the paramagnetic DOS showing high values for $x>0.75$
and with no $C$ $2p$ states at or near $E_{F}$, a possibility of
electronic distortion leading to a ferromagnetic ground state can
well be anticipated. Taking the cue, spin polarized calculations were
carried out for $MgC(Ni_{1-x}Co_{x})_{3}$ alloys for $0.7\leq x\leq1.0$. 

\begin{figure}[ht]
~

~

~

\includegraphics[%
  scale=0.3]{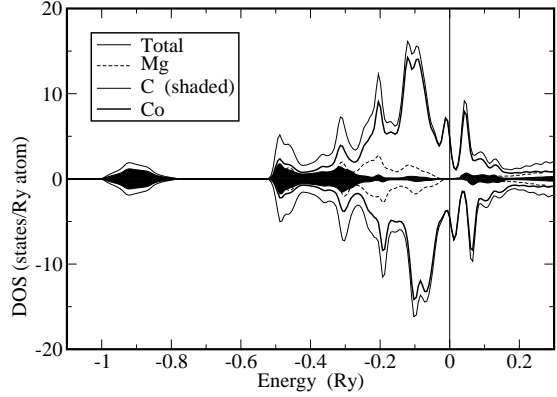}

\caption{\label{mgcco3-dos}The spin polarized total and sub-lattice resolved
partial spin up and down DOS of $MgCCo_{3}$. 
The vertical line through the energy zero represents the Fermi energy
and + and - DOS represent the majority and minority spin DOS respectively. }
\end{figure}

\begin{figure}[ht]
~

~

~

\includegraphics[%
  scale=0.3]{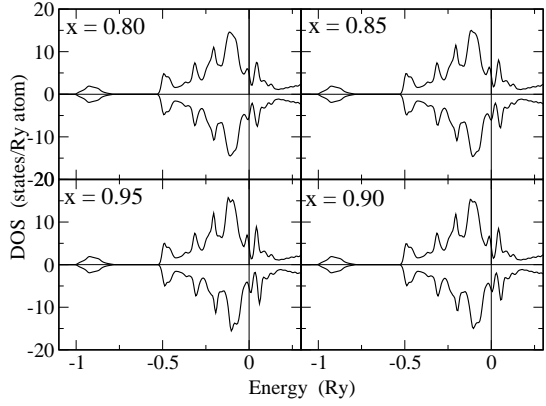}

\caption{\label{fig-co3-poltotDOS} The spin polarized total density of states
of $MgC(Ni_{1-x}Co_{x})_{3}$ alloys for $0.85\leq x\leq0.95$. The
vertical line through the energy zero represents the Fermi energy
and + and - DOS represent the majority and minority spin DOS respectively. }
\end{figure}

Fig. \ref{mgcco3-dos} shows the spin polarized, total and sub-lattice
resolved partial DOS of $MgCCo_{3}$. The deep states are primarily
$C$ $2s$ states. Higher in energy and close to $E_{F}$, the states
are predominately $Co$ $3d$ in character. As expected, the $Co$
$3d$ bands are exchange split due to magnetic effects. As a measure
of the exchange splitting on the energy scale, we find the difference
between the centres of the $3d$ up bands $C_{B}^{\uparrow}$ and
that of the down bands $C_{B}^{\downarrow}$ as $22$ mRy, which corresponds
to a local magnetic moment of $0.33$ $\mu_{B}$. These values are
consistent with earlier full-potential calculations. Small exchange
splitting of the $Mg$ and $C$ $2p$ bands are seen due to hybridization
effects, however, the splittings are negligible when compared to that
of the $Co$ $3d$ bands. 

The change in the DOS as a function of $x$ in $MgC(Ni_{1-x}Co_{x})_{3}$
alloys are shown in Fig.\ref{fig-co3-poltotDOS}. For both majority
and minority bands, the deep states essentially consist of $C$ $2s$
states. Higher in energy towards $E_{F}$, a gap opens up followed
by an admixture of $C$ $2p$ - metal $3d$ states. The spin up bands
are fully occupied and the $E_{F}$ for these states are pinned in
the pseudo-gap. For the spin down states, the $E_{F}$ moves slightly
inward with increasing $x$. Magnetism is thus governed by the change
in the position of $E_{F}$ in the minority DOS with respect to that
of the majority bands.

\begin{figure}[ht]
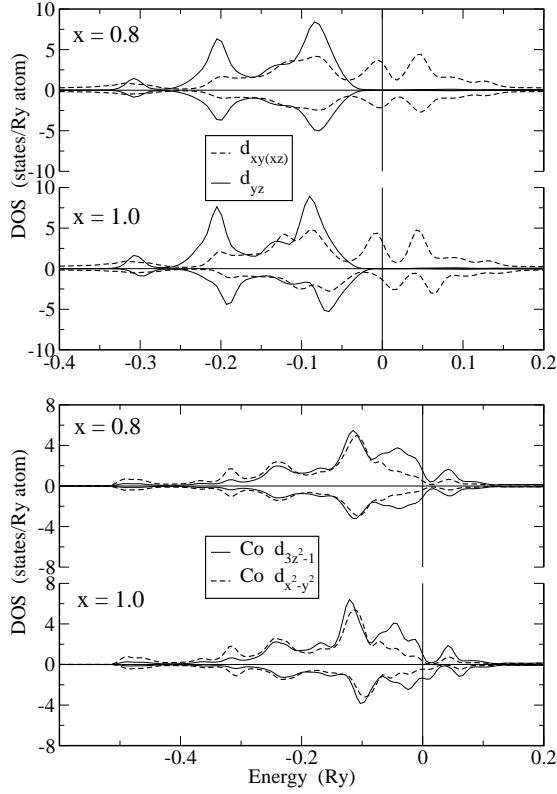

~

\includegraphics[%
  scale=0.3]{fig13a.eps}

\includegraphics[%
  scale=0.3]{fig13b.eps}

\caption{\label{fig-co3-pol-d-decom} The spin polarized, sub-lattice resolved
and orbital decomposed $Co$ $3d$ partial density of states for $0.85\leq x\leq0.95$
alloys. The vertical line through the energy zero represents the Fermi
energy. }
\end{figure}

The sub-lattice resolved, concentration weighted, $l-$decomposed
$3d$ DOS of $MgC(Ni_{1-x}Co_{x})_{3}$ alloys are as shown in 
Fig.\ref{fig-co3-pol-d-decom},
where the exchange splitting as well as the distortions can be seen
clearly. It also becomes evident that not all $d$ sub-states are
equally exchange split. Relatively, $d_{yz}$ and $d_{3z^{2}-1}$
bands show the maximum exchange splitting, while the states contained
in the $x-y$ plane, the distortions in the majority and minority
bands are more noticeable. Distortions result due to significant charge
transfer. Bonding is anisotropic due to covalent character along the
$x-y$ plane and rather metallic character along the $z$ direction.

\subsection{Magnetic properties}

\subsubsection{Variation of $Co$ local moment}

In general, when a paramagnetic DOS shows a high $N(E_{F})$, with
$E_{F}$ in the anti-bonding region, the system is close to a magnetic
instability\@. Under these conditions, distortion inevitably decreases
the total energy. The first choice, however, is a reshuffling of the
electronic states instead of the atoms getting displaced. The electrons
prefer to rearrange themselves resulting in a spontaneous magnetization
thereby lowering the electronic symmetry and annihilating the anti-bonding
states. 

According to Dronskowski $et$ $al$ \cite{intjqc-96-89,Z.Anorg-628},
anti-ferromagnetism is likely to step in when the paramagnetic DOS
has $E_{F}$ in a non-bonding region. For those systems, where $E_{F}$
is in the anti-bonding region, exchange splitting of bands towards
ferromagnetism is more likely. In $MgCNi_{3}$ the $E_{F}$ is in
the anti-bonding region, and hence it is closer to ferromagnetic instability
rather than an anti-ferromagnetic one. Further, anti-ferromagnetism
had been ruled out in $MgCNi_{3}$ owing to the absence of nesting
features in the Fermi surface topology.

\begin{figure}[ht]
~

~

~

\includegraphics[%
  scale=0.3]{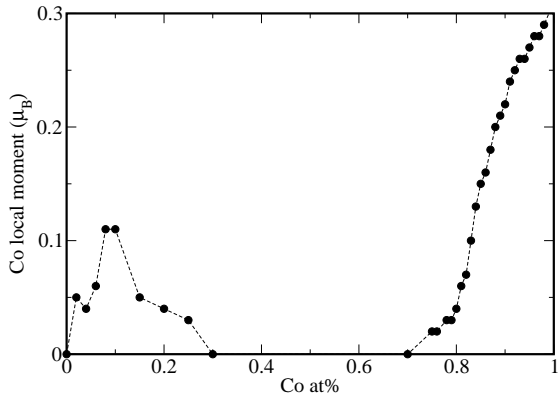}

\caption{\label{fig-co3-LocMom}The variation in the $Co$ local magnetic
moment in $MgC(Ni_{1-x}Co_{x})_{3}$ alloys calculated using the KKR-ASA-
CPA method in LDA for $0\leq x \leq 1$. }
\end{figure}

The present calculations find small moments at $Co$ site for $x\leq0.3$, as 
shown in Fig. \ref{fig-co3-LocMom}.
The results are inconsistent with experiments. Some of the factors
that can influence the magnetic moment at the $Co$ site are the choice
of the sphere radii, inadequate basis set used to expand the wave
functions, lack of lattice relaxation and local environment effects.
In this direction, calculations were carried out for different sphere
radii, but the local moment at the $Co$ site remained inevitable.
To include local environment effects, super-cell calculations based
on the LMTO-ASA as well as locally self consistent Green's function
method were employed, however the $Co$ remained magnetic. Since common
to all the above calculations was LDA, the appearance of local magnetic
moment at $Co$ site was thought to be due to the limitations of LDA
itself. 

It is known that first-principles LDA based calculations for materials
such as $FeAl$ \cite{prb-67-153106}, $Ni_{3}Ga$ \cite{prl-92-147201},
$Sr_{3}Ru_{2}O_{7}$ \cite{prb-63-165101}, $SrRhO_{3}$ \cite{prb-67-054507},
and $Na_{0.5}CoO_{2}$\cite{prb-61-13397,prb-68-020503} yield a finite
local moment in disagreement with the experiments which find them
to be paramagnetic. The error that evolved due to LDA in the present
self-consistent calculations for $MgC(Ni_{1-x}Co_{x})_{3}$ alloys,
was however corrected by means of the fixed-spin moment method, the
results of which we show below.

\subsubsection{On-site magnetic exchange interactions}

The total exchange coupling parameter $J_{0}$ is calculated as a
function of $x$, $x>0.7$ following the prescription of Liechtenstein
$et$ $al$ \cite{jpf-14-l125}. Essentially the method is based on
mapping the change in energy due to the deviation of a single spin
at a reference site {}``$o$'', from the collinear ferromagnetic
ground state onto an effective Heisenberg model. The change in the
energy corresponding to this small spin density perturbation is approximated
by the change in the sum of one electron energies by appealing to
Andersen's local force theorem. Using Lloyd's formula one can express
the sum of one electron energies in terms of scattering path operator
or the auxiliary Green's function. The latter can be evaluated from
the knowledge of potential functions and the structure constant matrix.
Mapping the change in energy onto a Heisenberg model results in an
expression for the exchange coupling constant:

\[
J_{0}=\sum_{i}J_{0i}\]
which is the sum of exchange interactions between the reference spin
and all its neighbors. In a mean field theory this coupling constant
is proportional to the Curie temperature $T_{C}$ of the system. For
a multi-component random alloy the mean field estimate of the $T_{C}$
can be assumed to be the concentration-weighted average of the coupling
constants calculated for the component atoms. Thus, for the case of
$MgC(Ni_{1-x}Co_{x})_{3}$ alloys, the average coupling constant can
be calculated as

\[
J_{0}\equiv(1-x)\,\, J_{Ni}+(x)\,\,J_{Co}\]
where $J_{i}$ ($i=Ni$ or $Co$) is given as

\[
J_{i}=-\frac{1}{4\pi}\sum_{L}\int^{E_{F}}\,\, dE\,\, Im\,\left\{ \Delta_{L}^{i}(z)\left[T_{L\uparrow}^{i}(z)-T_{L\downarrow}^{i}(z)\right]+\Delta_{L}^{i}(z)T_{L\uparrow}^{i}(z)\Delta_{L}^{i}(z)T_{L\downarrow}^{i}(z)\right\} \]
with

\[
\Delta_{L}^{i}=P_{L\uparrow}^{i}(z)-P_{L\downarrow}^{i}(z)\]

\[
P_{L\sigma}^{i}(z)=\left[z-C_{L\sigma}^{i}\right]\left[\Delta_{L\sigma}^{i}+\gamma_{L\sigma}^{i}\left(z-C_{L\sigma}^{i}\right)\right]^{-1}\]

\[
T_{L\sigma}^{i}(z)=\left\langle g_{L\sigma}(z)\right\rangle \left\{ 1+(P_{L\sigma}(z)-\tilde{P}_{L\sigma}(z)\left)\langle g_{L\sigma}(z)\right\rangle \right\} \]
where $\sigma$ is the spin index ($\uparrow$ or $\downarrow$),
and $P_{L\sigma}^{i}$ is the potential function of the component
$i$ for the orbital $L$ and spin $\sigma$. The potential function
has been expressed above in terms of $C$, $\Delta$ and $\gamma$
of the LMTO Hamiltonian, $z$ is the complex energy and $E_{F}$ is
the Fermi energy of the alloy, $\left\langle g_{L\sigma}(z)\right\rangle $
is the configurationally averaged auxiliary Green's function within
CPA and $\tilde{P}_{L\sigma}(z)$ is the coherent-potential of the
medium. The CPA calculations are performed by invoking site-site approximation.

\begin{figure}[ht]
~

~

~

\includegraphics[%
  scale=0.3]{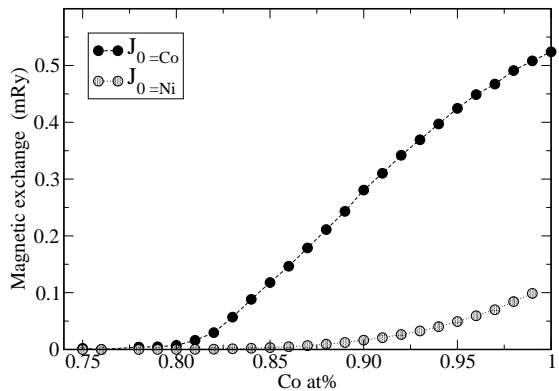}

\caption{\label{fig-co3ExcJ}The variation of the on-site exchange coupling
constant $J_{0}$ in mRy, as a function of $x>0.75$ in $MgC(Ni_{1-x}Co_{x})_{3}$
alloys. }
\end{figure}

Fig.\ref{fig-co3ExcJ} shows the variation in the $J_{Ni}$ and $J_{Co}$
as a function of $x$ in $MgC(Ni_{1-x}Co_{x})_{3}$ alloys. Both $J_{Co}$
and $\mu_{Co}$ (ref. Fig.\ref{fig-co3-LocMom}) follow a similar
trend. Both show a slow variation in and around the ferromagnetic
onset followed by a sharp increase, further getting saturated when
$x$$\rightarrow$$1$. Except for an overestimation of the exchange
coupling constant, the calculations are expected to produce the correct
trend. The disappearance of the average $Co$ local moment is accompanied
by disappearance of the average exchange interaction $J_{Co}$, a
result which can be related to the strong magneto-volume effect.

\subsubsection{Fixed-spin moment calculations}

The density functional theory is an exact ground state theory. However,
common approximations like LDA to the density functional theory is
thought to miss certain important physics due to the assumption of
a uniform electron gas especially when spin-fluctuation effects are
important. The effects of spin-fluctuations are, however, described
on a phenomenological level using the Ginzburg-Landau theory. Although
a robust quantitative theory based on first-principles is yet to be
implemented, it is possible to make an estimate of spin-fluctuation
effects based on LDA fixed-spin moment calculations.

According to Mazin and others \cite{nato-2003}, the overestimation
of the tendency of metals towards ferromagnetism within the LDA can
be used as an indicator of critical fluctuations in a material. However,
for this to be an effective indicator, competing states, like anti-ferromagnetism
need to be ruled out in each material. As mentioned above, since $E_{F}$
resides in the anti-bonding region as well as the absence of nesting
features in the Fermi surface topology of $MgCNi_{3}$ shows that
the material under consideration may be far from any anti-ferromagnetic
instability. 

For $MgC(Ni_{1-x}Co_{x})_{3}$, the overestimation of the moments
for $x<0.3$ could be taken as an indication of spin-fluctuations,
following the works of Singh and Mazin \cite{prb-64-140507}. However,
no attempt is made to determine the magnitude of the fluctuations
in the present work. The study is limited to understanding the propensity
of magnetism with $Co$ substitutions, for which one can use the phenomenological
Ginzburg- Landau coefficients, the input to which follows from the
fixed-spin moment method. 

\begin{figure}[ht]
~

~

~

\includegraphics[%
  scale=0.3]{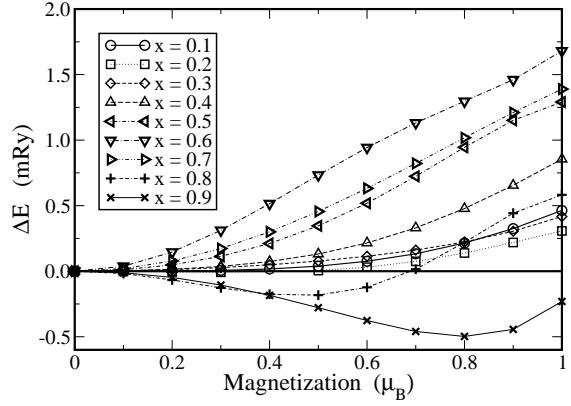}

\caption{\label{fig-co3-deltaEM}The variation in the $\Delta E(M)$ as a
function of magnetization $M$, calculated at the respective equilibrium
volume for $MgC(Ni_{1-x}Co_{x})_{3}$ alloys as a function of $x$,
as indicated in the figure. }
\end{figure}

Numerical calculations of magnetic energy $\Delta E(M)$ for $MgC(Ni_{1-x}Co_{x})_{3}$
alloys in the range $0\leq x\leq1$ are carried out by the fixed-spin
moment method. One can then write down a Ginzburg-Landau expansion
for the magnetic energy, $\Delta E(M)=\sum_{i\geq1}^n\frac{1}{2i}a_{2i}M^{2i}$
for $n=3$. The variation of $\Delta E(M)$ with respect to $M$ is
shown in Fig.\ref{fig-co3-deltaEM}. For $x\leq0.3$, the curves are
relatively flat near $M=0$ and fit well to the form as given above
with $n=3$. The sensitivity of $\Delta E(M)$ with $M$ becomes more
distinguishable as $x$ increases from $0.3$ and beyond, i.e., the
flatness of the curve disappears. For $x>0.7$ the magnetic energy
is negative, indicating a lower energy ferromagnetic state to be stable. 

\begin{figure}[ht]
~

~

~

\includegraphics[%
  scale=0.3]{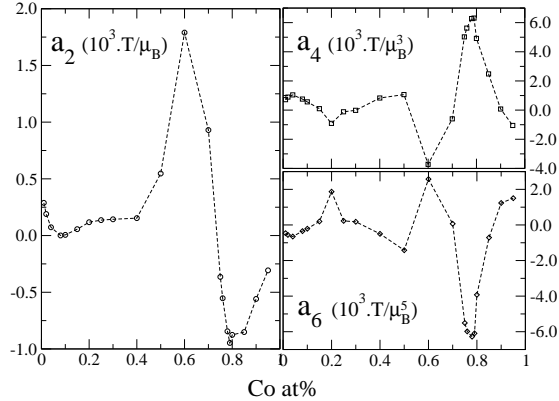}

\caption{\label{fig-Co3GinzLanCoeff}The variation of Ginzburg-Landau coefficients
as a function of $Co$ concentration $(x)$ in $MgC(Ni_{1-x}Co_{x})_{3}$
alloys.}
\end{figure}

Fig \ref{fig-Co3GinzLanCoeff} shows the variation of Ginzburg-Landau
coefficients $a_{2}$ and $a_{4}$ as a function of $x$ in $MgC(Ni_{1-x}Co_{x})_{3}$
alloys. The coefficient $a_{2}$ is of prime importance and is used
as a precursor to indicate the magnetic phase of the system. A positive
$a_{2}$ represents the paramagnetic state while a negative $a_{2}$
indicates the ferromagnetic state. From Fig \ref{fig-Co3GinzLanCoeff}
we find that initially, as $x$ increases, $a_{2}$ tends to zero,
but remains positive. This indicates that a ferromagnetic state is
unstable for low $Co$ rich alloys. However for $x>0.75$, $a_{2}$
is negative indicating a ferromagnetic state for the system. These
calculated results are now consistent with the experiments, which
show a definite paramagnetic phase for low $Co$ rich alloys and a
magnetic state for high $Co$ rich alloys.

As the unit cell volume is increased, $E_{F}$ is lowered in energy.
This will then sharpen the structures in the DOS. If $N(E_{F})$ then
satisfies the Stoner criterion, i.e., $I*N(E_{F})>1$, where $I$
is the Stoner exchange integral, then the system would become magnetic.
Having seen that the self-consistent calculations yield ambiguous
magnetic properties, one can extend the fixed-spin moment calculations
to the magnetization dependence of $MgC(Ni_{1-x}Co_{x})_{3}$ with
respect to the unit cell volume. 

\begin{figure}[!ht]
~

\includegraphics[%
  scale=0.3]{fig18a.eps}

\includegraphics[%
  scale=0.3]{fig18b.eps}

\includegraphics[%
  scale=0.3]{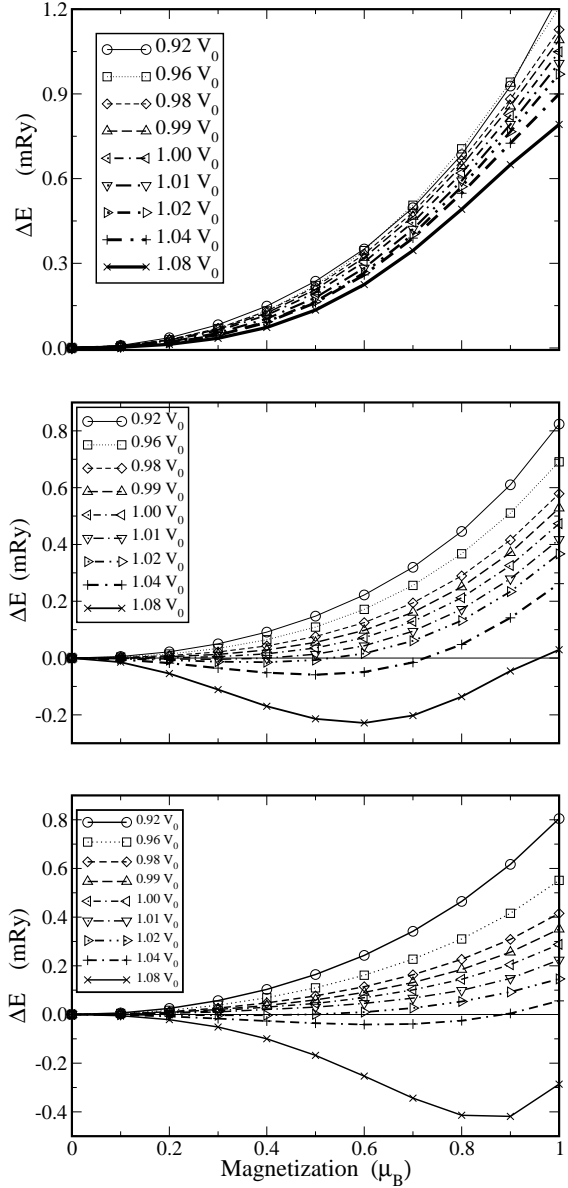}

\caption{\label{co3-fsm-01-DE}The variation of $\Delta E(M)$ as a function
of $M$ for $MgC(Ni_{1-x}Co_{x})_{3}$ alloys at various
volume ratios as indicated in the figure for (a) $x=0.01$ (upper panel), 
(b) $x=0.1$ (middle panel) and (c) $x=0.2$ (lower panel). }
\end{figure}

Fig.\ref{co3-fsm-01-DE} shows the variation of
the magnetic energy as a function of volume, expressed in terms of its 
equilibrium volume  $V_0$, 
for $x=0.01, 0.1$ and $0.2$, respectively. 
For $x=0.01$, an expansion by $8$\%
shows no magnetic phase for the alloy. However, for both $x=0.1$
as well as $x=0.2$, a magnetic phase looks possible just above its
equilibrium volume. The curves dip into the negative region of the
magnetic energy as can be seen from the middle and the bottom panel of Fig. \ref{co3-fsm-01-DE},
respectively, where the ferromagnetic state
is lower in energy.

For $x=0.1$ and $0.2$, a ferromagnetic phase becomes stable just
above its equilibrium volume suggesting a magneto-volume instability
in $MgC(Ni_{1-x}Co_{x})_{3}$ alloys. Fitting the above data to the
Ginzburg-Landau expression for $n=3$ for the three concentrations
$x=0.01$, $0.10$ and $0.20$, one finds that for expanded volumes
a magnetic phase becomes definitely possible as $Co$ $at$\% increases
in $MgC(Ni_{1-x}Co_{x})_{3}$. 

\begin{figure}[!ht]
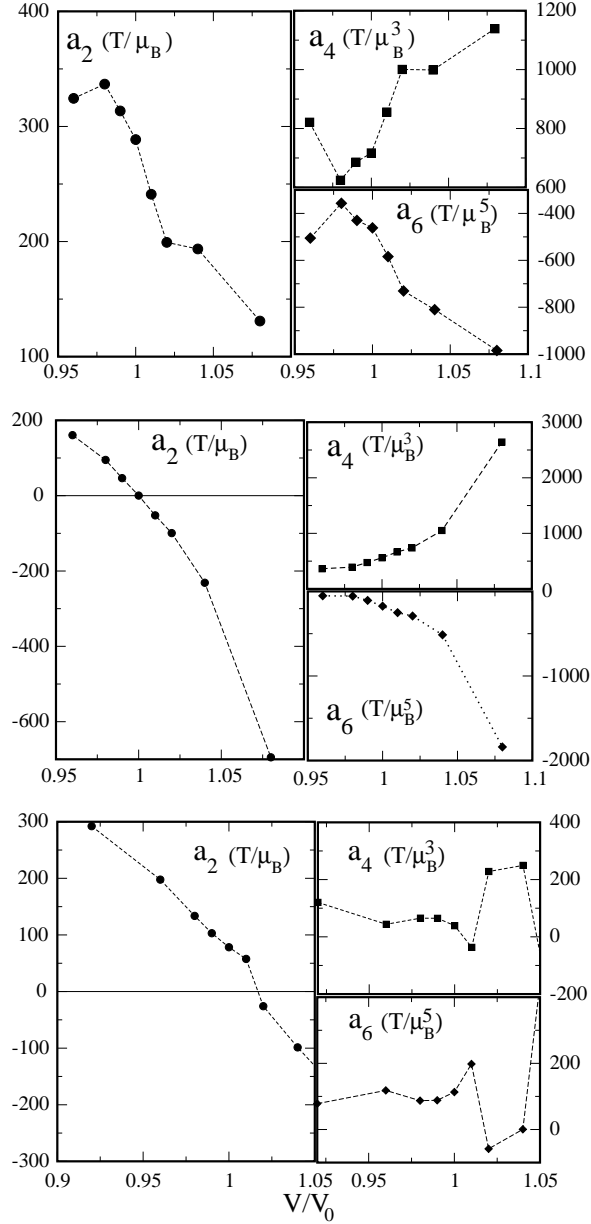

~

~

~

\includegraphics[%
  scale=0.3]{fig19a.eps}

\includegraphics[%
  scale=0.3]{fig19b.eps}

\includegraphics[%
  scale=0.3]{fig19c.eps}

\caption{\label{co3-fsm-01-GLc}The variation of the Ginzburg-Landau coefficients,
$a_{2}$(in units of $\frac{T}{\mu_{B}}$), $a_{4}$(in units of $\frac{T}{\mu_{B}^{3}}$),
and $a_{6}$ (in units of $\frac{T}{\mu_{B}^{5}}$) as a function
of volume for $MgC(Ni_{1-x}Co_{x})_{3}$ alloys for (a) $x=0.01$ (upper panel), 
(b) $x=0.1$ (middle panel) and (c) $x=0.2$ (lower panel). }
\end{figure}

Fig.\ref{co3-fsm-01-GLc} shows the variation
of the Ginzburg- Landau coefficients as a function of volume ratio
for $x=0.01$, $0.10$ and $0.20$, respectively. For $\frac{V}{V_{0}}$
slightly greater than unity, for both $x=0.1$ and $x=0.2$, the coefficient
$a_{2}$ reverses its sign, indicating a phase transition with respect
to volume. This confirms that these alloys are on the verge of a magneto-volume
instability. The higher order coefficients play a significant role
in determining the alloy magnetic properties. 

It has also been noted that the accuracy of the total energy, evaluated
by means of ASA is limited to a few mRy. From calculations, one finds
that the change in the energy is far too small for any reliable quantitative
analysis. However, within the approximations mentioned and with well
defined $\mathbf{k}-$mesh in the Brillouin zone, one can certainly
infer that the trend produced via the fixed-spin moment method for
$MgC(Ni_{1-x}Co_{x})_{3}$ alloys are qualitatively correct.

\section{Conclusions}

For low $Co$ rich $MgC(Ni_{1-x}Co_{x})_{3}$ alloys, we find the
possibility of enhanced spin-fluctuations in the material, which is
close to a ferromagnetic instability. However, for $Co$ rich alloys,
a definite ferromagnetic ground state exists. At the magnetic cross-over,
i.e., for $x=0.75$, a concentration independent variation in the
structural properties are determined. For example, the lattice constant
and the bulk modulus appear to be constant under the given approximations
of the KKR-ASA-CPA theory. The pressure derivative of the bulk modulus
indicates, within the Debye approximation, a significant change in
the averaged vibrational modes. Our calculations show that the electronic
structure of the disordered $MgC(Ni_{1-x}Co_{x})_{3}$ alloys deviates
significantly from that of the rigid band model. The striking feature
is the recession of $C$ $2p$ states towards lower energies as a
function of increasing $x$ in $MgC(Ni_{1-x}Co_{x})_{3}$ alloys.
The self-consistent calculations overestimate the magnetic moments
at the $Co$ site for low $Co$ rich alloys, due to the limitations
in the local-density approximation. Corrections to the overestimation
in the magnetic moments are accomplished by means of fixed-spin moment
method, in conjunction with the Ginzburg-Landau free energy functional.
It then follows from the fixed-spin moment calculations that for $x<0.7$
the material is definitely paramagnetic.

\end{document}